\documentclass[aps,prd,superscriptaddress,amsmath,amssymb,nofootinbib,nobibnotes,twocolumn,preprintnumbers]{revtex4-2}
\usepackage{graphicx}
\usepackage{dcolumn}
\usepackage{amsmath}
\usepackage{float}
\usepackage{makecell}
\usepackage{bm}
\usepackage{comment}
\usepackage{subcaption}
\usepackage{slashed}
\usepackage{gensymb}
\usepackage{tikz}
\usepackage[normalem]{ulem}
\usepackage[compat=1.1.0]{tikz-feynman}
\usepackage{ragged2e}
\usepackage[utf8]{inputenc}
\usepackage{tabularx}
\usepackage{array}
\usepackage{graphics}
\graphicspath{{}}
\usepackage{psfrag}
\usepackage{epsfig}
\usepackage{amssymb}
\usepackage{setspace}
\usepackage{rotating}
\usepackage{colortbl}
\usepackage{slashed}
\usepackage{enumitem}
\usepackage[T1]{fontenc}
\makeatletter
\usepackage{textcomp}
\usepackage[normalem]{ulem}
\usepackage[colorlinks=true,allcolors=blue]{hyperref}
\usepackage[capitalize]{cleveref}
\usepackage{accents}

\definecolor{lightblue}{rgb}{0.68, 0.85, 0.9}
\definecolor{deepblue}{rgb}{0.0, 0.0, 0.55}

\def\be{\begin{equation}}
\def\ee{\end{equation}}
\def\bea{\begin{eqnarray}}
\def\eea{\end{eqnarray}}

\definecolor{nicegreen}{rgb}{0., 0.75, 0.46}

\begin{document}

\begin{flushright}
MI-HET-883
\end{flushright}

\title{Producing the GeV Galactic Center Excess via Cosmic Ray-Dark Matter Scattering}

\author{Bhaskar Dutta}
\email{dutta@tamu.edu}
\affiliation{
Mitchell Institute for Fundamental Physics and Astronomy,
Department of Physics and Astronomy, Texas A\&M University, College Station, TX 77843, USA
}
\author{Debopam Goswami}
\email{debopam22@tamu.edu}
\affiliation{
Mitchell Institute for Fundamental Physics and Astronomy,
Department of Physics and Astronomy, Texas A\&M University, College Station, TX 77843, USA
}
\author{Jason Kumar}
\email{jkumar@hawaii.edu}
\affiliation{
Department of Physics \& Astronomy, University of Hawai‘i, Honolulu, HI 96822, USA
}
\author{Mudit Rai}
\email{muditrai@tamu.edu}
\affiliation{
Mitchell Institute for Fundamental Physics and Astronomy,
Department of Physics and Astronomy, Texas A\&M University, College Station, TX 77843, USA
}
\author{Deepak Sathyan}
\email{dsathyan@tamu.edu}
\affiliation{
Mitchell Institute for Fundamental Physics and Astronomy,
Department of Physics and Astronomy, Texas A\&M University, College Station, TX 77843, USA
}

\begin{abstract}
In this work, we propose a novel mechanism for generating gamma rays from the Galactic Center via scattering of cosmic-ray protons off dark matter in the Milky Way halo, in contrast to conventional explanations based on dark matter annihilation.
We present two examples of this framework that produce an observable photon signal.
In the inelastic dark matter model, cosmic rays up-scatter a lighter dark matter particle, with the subsequent decay of the heavier particle yielding two photons.
In the elastic dark matter model, an energetic photon is directly produced in the final state of a 2-to-3 scattering process.
We show that, for a range of viable model parameters, this framework provides a fit to the observed Galactic Center gamma-ray excess spectrum comparable to those obtained from dark matter annihilation and millisecond pulsar models. 
Our results open a new avenue for interpreting gamma-ray observations of the Galactic Center.
\end{abstract}

\maketitle

\textbf{Introduction}---The Galactic Center excess (GCE)~\cite{Goodenough:2009gk, Hooper:2010mq, Hooper:2011ti, Morselli:2010ty, Abazajian:2012pn, Fermi-LAT:2015sau} of GeV-energy gamma rays observed by the \textit{Fermi} Large Area Telescope (\textit{Fermi}-LAT) has inspired many interesting ideas regarding its source. Proposed production mechanisms of these $\gamma$-rays include annihilation of WIMP dark matter (DM) \cite{Goodenough:2009gk, Calore:2014xka, Gordon:2013vta, Hooper:2013rwa, Queiroz:2026tah}, millisecond pulsars (MSPs)~\cite{Abazajian:2010zy, Fermi-LAT:2017opo, Bartels:2015aea}, 
cosmic-ray outbursts~\cite{Carlson:2014cwa, Petrovic:2014uda, Cholis:2015dea},
and additional sources of $\gamma$-rays from the supermassive black hole at the Galactic Center~\cite{Carlson:2014cwa,Carlson:2015ona,Carlson:2016iis}. 

These potential sources of the GCE, while consistent with \textit{Fermi}-LAT observations, also yield other predictions which may be in tension with complementary observations. For example, in the case of annihilating dark matter, emission of $\gamma$-rays from dwarf spheroidal galaxies (dSphs) may also be observed (see, for example,~\cite{Fermi-LAT:2016uux, Abazajian:2015raa}).
Cosmic ray-proton scattering arising from hadronic cosmic-ray outbursts typically produces photons along the galactic disk~\cite{Carlson:2014cwa, Cholis:2015dea}
and can be constrained by observations of photon emission from the galactic disk~\cite{Cholis:2015dea}. 

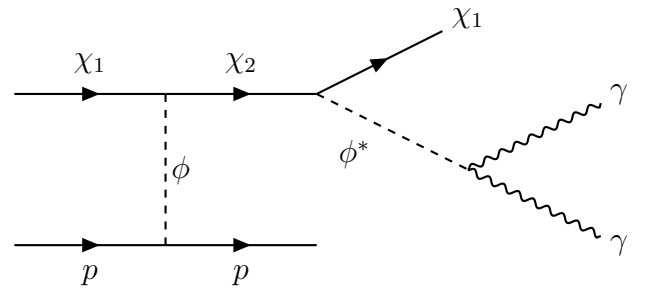
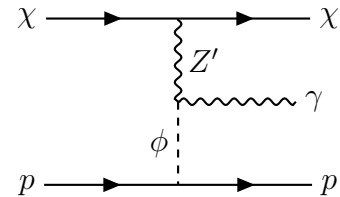
\begin{figure}[!htbp]
    \centering
    \begin{subfigure}{0.5\textwidth}
        \centering
        \begin{tikzpicture}
        \tikzset{every node/.style={font=\large}}  
            \begin{feynman}
                \vertex (a) at (0,0);
                \vertex (a1) at (1,0.4) {$\chi_1$};
                \vertex (b) at (2,0);
                \vertex (c) at (4,0);
                \vertex (c1) at (3.0,0.4) {$\chi_2$};
                \vertex (d) at (0,-2);
                \vertex (d1) at (1,-2.4) {$p$};
                \vertex (e) at (2,-2);
                \vertex (e1) at (2.2,-1) {$\phi$};
                \vertex (f) at (4,-2);
                \vertex (f1) at (3,-2.4) {$p$};
                \vertex (g) at (6,1) {$\chi_1$};
                \vertex (h) at (6,-1);
                \vertex (i) at (4.5,-0.8) {$\phi^*$};
                \vertex (j) at (8,0) {$\gamma$};
                \vertex (k) at (8,-2) {$\gamma$};
                \diagram*{
                (a) -- [fermion, thick] (b) -- [fermion, thick](c) -- (c) -- [fermion, thick](g),
                (b) -- [scalar, thick](e),
                (d) -- [fermion, thick](e) -- [fermion, thick](f),
                (c) -- [scalar, thick](h),
                (h) -- [photon, thick](j), (h) -- [photon, thick](k)
                };
            \end{feynman}
        \end{tikzpicture}
        \captionsetup{justification=Justified, singlelinecheck=false}
        \caption{ Inelastic DM: up-scattering of the ambient DM ($\chi_1$) and subsequent decay of $\chi_2$ into $\chi_1$ and two photons}
        \label{subfig:InelasticScattering}
    \end{subfigure}

    \vspace{1.5em}
    \begin{subfigure}{0.5\textwidth}
        \centering
        \begin{tikzpicture}
            \begin{feynman}
                \tikzset{every node/.style={font=\large}}
                \vertex (i) at (-2, 1.1) {$\chi$};
                \vertex (f) at (2, 1.1) {$\chi$};
                \vertex (Ni) at (-2,-1.1) {$p$};
                \vertex (Nf) at (2,-1.1) {$p$};
                \vertex (if) at (0, 1.1);
                \vertex (Nif) at (0, -1.1);
                \vertex (c) at (0,0);
                \vertex (g) at (1.8,0) {$\gamma$};
                \diagram*{
                    (i) -- [fermion,  thick] (if) -- [fermion,  thick] (f),
                    (Ni) -- [fermion, thick] (Nif) -- [fermion, thick] (Nf),
                    (if) -- [boson,  thick, edge label=$Z^\prime$] (c),
                    (c) -- [scalar,  thick, edge label'=$\phi$] (Nif),
                    (c) -- [photon,  thick] (g),
                };
            \end{feynman}
        \end{tikzpicture}
        \captionsetup{justification=Justified, singlelinecheck=false}
        \caption{Elastic DM: scattering of a cosmic-ray proton with DM $\chi$ producing a photon via $Z'$ and $\phi$ exchange.}
        \label{subfig:elastic_scattering}
    \end{subfigure}
    \captionsetup{justification=Justified, singlelinecheck=false}
    \caption{Feynman diagrams of the CR-DM scattering processes with inelastic and elastic DM models.}
\end{figure}

\begin{figure*}[!htbp]
    \centering
    \includegraphics[width=0.8\textwidth]{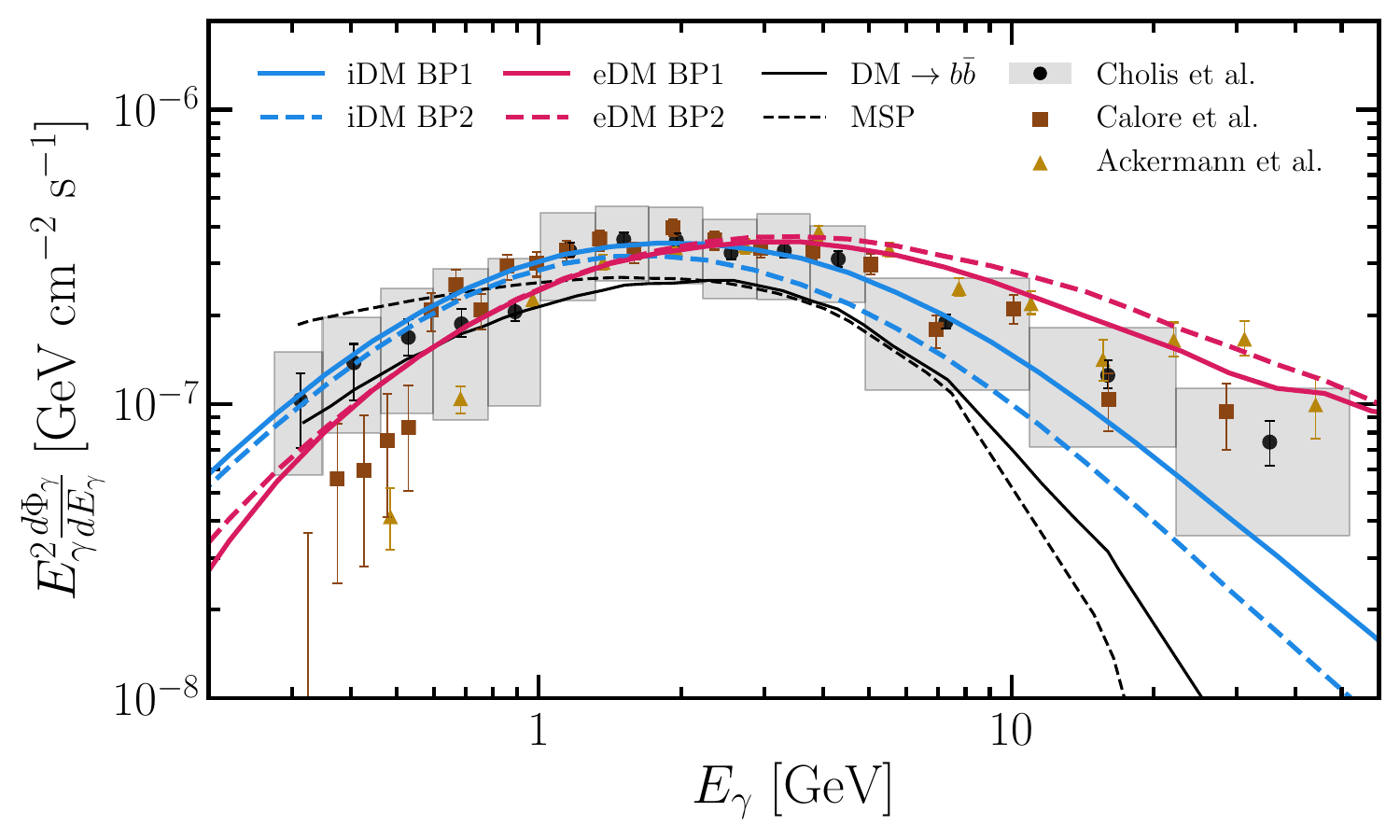}
    \captionsetup{justification=Justified, singlelinecheck=false}
    \caption{Spectra of our CR-DM models (colored lines) in comparison to the best fit spectra by DM annihilation to $b \overline{b}$ (solid black) and MSPs (dashed black) to the GCE obtained from  Cholis et al.~\cite{Cholis:2021rpp} (black circles), with associated statistical and correlated systematics (gray boxes) shown. 
    We show the spectra of two benchmark points from the inelastic DM model (blue solid \& dashed) and two benchmark points from the elastic DM model (red solid \& dashed).
    Additional background models by Calore et al.~\cite{Calore:2014xka} (maroon squares) and the Fermi-LAT collaboration~\cite{Fermi-LAT:2017opo} (gold triangles) are shown as examples of different GCE spectra, all featuring a peak between 1 and 5 GeV. 
    The goodness of fit of our CR-DM spectra, as well as the DM annihilation and MSP spectra, are shown in~\cref{tab:chiSq}.
    }
    \label{fig:moneyPlot}
\end{figure*}

In this \textit{Letter}, we propose a novel production mechanism for the GCE: hadronic cosmic ray-dark matter (CR-DM) scattering that produces high-energy photons. Such scatterings will dominantly produce photons where the DM density is largest, near the Galactic Center, with little emission from the rest of the galactic disk. The reduced dark matter density and decreased cosmic ray flux at the locations of dSphs would explain the absence of comparable $\gamma$-ray excesses observed in dwarf galaxies.

In this study, we consider two sub-GeV DM models: (1) an inelastic dark matter model~\cite{Tucker-Smith:2001myb, Bell:2021xff}, in which up-scattering by cosmic ray protons $\chi_1 + p \rightarrow \chi_2 + p$ and subsequent decay ($\chi_2 \rightarrow \chi_1 + 2\gamma$) produces two photons as shown in~\cref{subfig:InelasticScattering}, and (2) an elastic dark matter model involving scalar and vector mediators, in which a 2-to-3 process $\chi + p \rightarrow \chi + \gamma + p$ produces a high-energy photon~\cite{Dutta:2025fgz, Dutta:2025sba} as shown in~\cref{subfig:elastic_scattering}. We show that, for each model, there exist benchmark points in its parameter space that yield a $\gamma$-ray emission spectrum consistent with the best-fitting GCE models in~\cref{fig:moneyPlot}. 

CR–DM scattering has been considered in the literature~\cite{Bringmann:2018cvk,Ema:2018bih,Cappiello:2018hsu,Dent:2019krz} and explored as a source of photons in Ref.~\cite{Bloom:1997vm,Profumo:2011jt}. In Ref.~\cite{Profumo:2011jt}, the $s$-channel scattering of CR electrons or protons with WIMP DM, accompanied by the emission of an additional photon, was analyzed, and it was found that the resulting GeV photon energy flux is insufficient to account for the GCE. The key aspect of our solution lies in the sub-GeV DM, which enables CR protons to produce GeV-scale photons via $t$-channel scattering.

This \textit{Letter} is organized as follows: We briefly overview the modeling of the cosmic ray flux, followed by descriptions of the inelastic and elastic dark matter models and relevant constraints. Next, we compute the photon energy flux from CR-DM scattering. The results are presented in the final section, with comparisons to spectra from DM annihilation (DM $\to$ $b \overline{b}$) and MSPs.


\smallskip
\textbf{Cosmic Ray-Dark Matter Scattering}---To thoroughly describe CR-DM scattering and the resulting $\gamma$-ray signal, we first discuss the model of the galactic cosmic ray flux, followed by dark matter models considered. These are the two primary ingredients used to compute the spectrum of outgoing photons from the scattering processes considered.


We model the cosmic ray flux (CRF) in the Milky Way as a cylinder with a radius of 10 kpc\footnote{The radius of the CRF cylinder can be larger than 10 kpc. We choose 10 kpc as a conservative estimate.} and a half-height of 4 kpc, following Refs.~\cite{Strong:1998fr, Guo:2020oum}. This approximation is consistent with \texttt{GALPROP}~\cite{Strong:1998pw, Carlson:2014cwa}. The energy spectrum of the cosmic ray flux is obtained from local measurements~\cite{Anjos:2020sxt}. Modifying the CRF in the Galactic Center has been shown to affect the background modeling of the GCE~\cite{Carlson:2014cwa, Carlson:2015ona, Carlson:2016iis}, but we do not make such modifications. 
For the signal process we consider, we restrict attention to the proton component of the CRF. Additional contributions may arise from electron, positron, and heavy-ion fluxes, but they are not considered in this work.

The dominant contribution to the flux of $\sim$ GeV photons produced by CR-DM scattering arises from cosmic rays with energies $\lesssim 100$~GeV.
This is due in part to the power-law suppression of the cosmic ray flux at higher energies~\cite{Anjos:2020sxt} convolved with the differential cross section. We can therefore neglect the high-energy tail of the CRF, where uncertainties in the power-law scaling are larger.

\begin{table*}[!htbp]
    \centering
    \renewcommand{\arraystretch}{1.2}
    \begin{tabular}{ | c | c | c | c | c | c | c |}
        \hline
        \multicolumn{7}{|c|}{\textbf{Inelastic DM Model}} \\
        \hline
        BP & $m_{\chi_1}$ [GeV] & $\Delta$ [GeV] & $m_\phi$ [GeV] & $g_{12}$ & $g_{\phi pp}$ & $g_{\phi \gamma \gamma}$~[GeV$^{-1}$]\\
        \hline
        iDM BP1 & 0.08 & 0.24 & 0.25 & 0.80 & 0.04 & $1.0 \times 10^{-5}$\\
        iDM BP2 & 0.16 & 0.31 & 0.32 & 1.00 & 0.05 & $1.0 \times 10^{-5}$\\
        \hline

        \multicolumn{7}{|c|}{\textbf{Elastic DM Model}} \\
        \hline
        BP & $m_{\chi}$ [GeV] & $m_{Z^{'}}$ [GeV] & $m_\phi$ [GeV] & $\alpha_D$ 
        & \multicolumn{2}{c|}{$g_{\phi Z^{'} \gamma}g_{\phi pp}$~[GeV$^{-1}$]} \\
        \hline
        eDM BP1 & 1.00 & 0.19 & 0.20 & 0.16 & 
        \multicolumn{2}{c|}{1.96} \\
        eDM BP2 & 0.50 & 0.14 & 0.15 & 0.12 & 
        \multicolumn{2}{c|}{1.43} \\
        \hline

    \end{tabular}
    \captionsetup{justification=Justified, singlelinecheck=false}
    \caption{Benchmark points (BP) illustrating representative parameter choices for both models. For inelastic dark matter, $g_{12}$ denotes the off-diagonal coupling of the scalar mediator $\phi$ to $\chi_1$ and $\chi_2$. The mass splitting $\Delta = m_{\chi_2}-m_{\chi_1}$ is chosen to satisfy $\Delta < m_\phi$, so that $\chi_2$ undergoes the three-body decay. For the elastic dark matter model, $g_D=\sqrt{4\pi\alpha_D}$ is the coupling between the $Z'$ mediator and DM, and $m_{\rm DM} \sim$ GeV as motivated by asymmetric DM.}
    \label{tab:benchmarkpoints}
\end{table*}

Now, we describe two dark matter models: a model of inelastic dark matter with a scalar mediator, with a decay of the heavier DM component producing two photons, and a model of elastic 
dark matter with two mediators, with an outgoing photon produced by a 2-to-3 scattering process. We then discuss constraints on each DM model.

\smallskip
\textit{Inelastic Dark Matter (iDM)---} First, we consider an inelastic dark matter model~\cite{Tucker-Smith:2001myb, Izaguirre:2014dua, Giudice:2017zke, Dutta:2019fxn} with two species: the lighter, stable component $\chi_1$ and the heavier component $\chi_2$ with mass difference $\Delta$.
The interaction Lagrangian is given by 
\begin{equation}
\mathcal{L}_{\rm{iDM}} =
\, g_{12}\,\overline{\chi}_1 \chi_2  \phi 
+ g_{\phi pp}\,\phi\,\overline{p}p 
- \frac{1}{2} g_{\phi\gamma\gamma}\,\phi F_{\mu\nu}F^{\mu\nu} + \rm{h.c.} ,
\label{inelastic}\end{equation}
where $\phi$ is the scalar mediator between the dark sector and SM, $g_{12}$ is the off-diagonal coupling of DM particles with $\phi$ 
(for our benchmark cases, the diagonal coupling is negligible), $g_{\phi pp}$ is the effective coupling between $\phi$ and protons, and $g_{\phi \gamma \gamma}$ is the non-renormalizable coupling between $\phi$ and the SM photons.

Depending on the mass splitting $\Delta = m_{\chi_2} - m_{\chi_1}$ and the mediator mass $m_{\phi}$, the scalar $\phi$ produced from $\chi_2$ decay may be either on-shell or off-shell, followed by its decay into a photon pair.  
This analysis examines the process in which a CR proton up-scatters $\chi_1$ to $\chi_2$, followed by the three-body decay of $\chi_2$ that produces two photons via an intermediate off-shell $\phi^*$, as illustrated in~\cref{subfig:InelasticScattering}. 

We present two benchmark points (BPs) for the inelastic model, shown in~\cref{tab:benchmarkpoints}. Benchmark point 1 has 4 free parameters: $m_{\chi_1}$, $\Delta$, $m_\phi$, and the combination of couplings. However, for BP2, we fix $m_\phi = 2 m_{\chi_1}$.\footnote{Other choices of $m_\phi/m_{\chi_1}$ work as well, as in BP1.}
Benchmark points are selected to yield an energy spectrum with a peak near $\mathcal{O}(1)$ GeV. The lifetime of $\chi_2$ is sufficiently short on cosmological timescales, resulting in ambient dark matter composed entirely of $\chi_1$.

\smallskip
\textit{Elastic Dark Matter (eDM)---}
We consider another simplified framework with a light scalar $\phi$ and a light vector boson $Z^\prime$ that mediate interactions between the Standard Model and a single-component fermion dark matter particle $\chi$. The effective Lagrangian~\cite{Dutta:2025fgz} given by

\begin{equation}
\mathcal{L}_\mathrm{eDM} \supset g_D \overline{\chi}\gamma^\mu\chi  Z^\prime_\mu + y_\chi \overline{\chi}\chi \phi -\frac{1}{2}g_{\phi Z^\prime \gamma}\phi F_{\mu\nu} Z^{\prime \mu\nu}
+ g_{\phi pp}\phi\overline{p}p ,
\label{eq:lagrChi}
\end{equation}
where $m_\chi$ denotes the mass of the ambient DM $\chi$, $F_{\mu\nu}$ and $Z^\prime_{\mu\nu}$ are the photon and $Z^\prime$ field-strength tensors, and $g_{\phi p p}$ is the effective scalar proton coupling as discussed for the iDM model. 
The dimensionful coupling \( g_{\phi Z' \gamma} \) mediates photon emission, as shown in~\cref{subfig:elastic_scattering}. 
One motivation for considering such a DM model with a GeV-scale mass is asymmetric dark matter~\cite{Kaplan:2009ag}, which relates the baryon and DM asymmetries to explain the density ratio between baryons and DM. 

For the elastic DM benchmark points, in general, there are 4 free parameters: $m_\chi$, $m_{Z'}$, $m_\phi$, and the combination of couplings.
For the BPs shown in~\cref{tab:benchmarkpoints}, we fix the mass ratio of the mediators $m_{Z'} = 0.95 \, m_\phi$, leaving 3 free parameters for both BPs.

For both models, we use the dipole form factor for the $\phi-p-p$ vertex (details are provided in Appendix~\ref{FormFactor}).


\textit{Model constraints---}We first discuss constraints that apply to both dark matter models considered to produce GeV $\gamma$-rays. Then we describe the constraints that apply to each specific model. 

Scattering of low-mass ambient DM by cosmic rays produces a flux of boosted dark matter that can yield signals in direct detection experiments such as XENON1T~\cite{XENON:2018voc, XENON:2019zpr}, PandaX-4T~\cite{PandaX:2022xqx, PandaX:2023xgl, PandaX:2024muv}, and LZ~\cite{LZ:2022lsv, LZ:2018qzl, LZ:2024zvo}. In these lower-threshold detectors, we estimate $\mathcal{O}(1)$ signal events for a $1~\mathrm{ton}\cdot\mathrm{year}$ exposure, which falls below the observed limits once realistic detector efficiencies are taken into account. We find that our BPs are consistent with current direct-detection bounds using the scalar-proton coupling as shown in Eqs.~\ref{inelastic} and \ref{eq:lagrChi}. Furthermore, our iDM and eDM BPs are consistent with constraints from direct detection of ambient dark matter, given the local DM density.

The scalar proton coupling $g_{\phi p p}$ appearing in both iDM and eDM models is also consistent with constraints from NA62~\cite{NA62:2021bji}, PIENU~\cite{PIENU:2021clt},  MAMI~\cite{A2atMAMI:2014zdf}, mono-jet~\cite{Batell:2018fqo}, etc. (e.g., Ref.~\cite{Dutta:2025fgz}).

In addition to the above constraints, there are also constraints specific to iDM to consider. Laboratory searches predominantly constrain the mediator couplings to the SM, including $g_{\phi pp}$ and $g_{\phi\gamma\gamma}$. For mediator masses $m_\phi \simeq 180\text{--}250~\mathrm{MeV}$ and couplings $g_{\phi\gamma\gamma} \sim 10^{-5}~\mathrm{GeV}^{-1}$ 
the existing bounds from NA64~\cite{Banerjee:2019pds}, BaBar~\cite{BaBar:2017tiz}, DELPHI~\cite{DELPHI:2008uka}, etc.~are 
satisfied (e.g., Ref.~\cite{Dutta:2025sba}). Importantly, the exact value of $g_{\phi\gamma\gamma}$ has a negligible impact on the predicted photon flux unless $\chi_2$ becomes long-lived on scales comparable to the GCE region.
 
In the eDM scenario, the dimension-5 operator $\phi F_{\mu\nu} Z^{\prime \mu\nu}$ is constrained by fixed-target and beam-dump searches for light states, as well as by indirect probes such as electron and muon $g-2$~\cite{deNiverville:2018hrc}. Our choice of mediator masses (i.e., $m_{Z'} \sim m_{\phi}$) allows these constraints to be satisfied, although residual theoretical uncertainties remain. In particular, searches at BaBar~\cite{BaBar:2001yhh} are largely insensitive to this operator when the mass splitting $\Delta m = m_\phi - m_{Z'}$ is small, as it leads to soft photon spectra~\cite{Dutta:2025fgz}.
Additionally, observations of the Bullet Cluster place a limit on the strength of self-interacting DM~\cite{Markevitch:2003at, Harvey:2015hha, Robertson:2016xjh} through the vector mediator in our model. The considered eDM benchmark points are safe from this constraint.

Since the mediator masses we consider are $\gtrsim 200~\mathrm{MeV}$, astrophysical constraints from supernova cooling (e.g., SN1987A) are not applicable~\cite{Dev:2020eam, Chang:2018rso,Croon:2020lrf}. In any case, such bounds are highly model- and medium-dependent, as production and trapping rates depend sensitively on in-medium effects and additional interactions~\cite{Chang:2018rso, Sung:2021swd, Hardy:2016kme, Hook:2021ous}.


\smallskip
\textbf{Photon Flux from CR-DM scattering}---
The differential photon flux integrated over the $40^\circ \times 40^\circ$ region of interest (ROI), excluding the inner $2^\circ$ above and below the galactic plane~\cite{Hu:2025thq} is given by
\begin{equation}
\frac{d \Phi_\gamma}{dE_\gamma} = 
\frac{D_{\rm ROI}}{4\pi m_{\rm DM}}
\int dE_p \, \frac{d\Phi_p}{dE_p}
\frac{d\sigma(E_p, E_\gamma)}{dE_\gamma},
\label{eq:flux_calculation}
\end{equation}
where $D_{\rm ROI} \equiv \int_{\Delta \Omega} d\Omega \int d\ell ~ \rho_{\rm DM}(r) $ is the total $D$-factor (decay-like $J$-factor) integrated over the ROI and the line-of-sight, $d\sigma(E_p)/dE_\gamma$ denotes the differential cross section of CR-DM scattering with respect to the photon energy $E_\gamma$, for an incoming CR proton of energy $E_p$, and $d\Phi_p/dE_p$ is the CR proton flux integrated over the solid angle $4\pi$.

In this analysis, $\rho_{\rm DM}(r)$ is considered to be the generalized Navarro-Frenk-White (gNFW) profile~\cite{Navarro:1996gj} (with $\gamma$=1.25~\cite{Hu:2025thq}) given by
\begin{align}\label{DMprofile}
\rho_{\rm gNFW}(r) = \rho_s \Bigg(\frac{r}{r_s}\Bigg)^{-\gamma}\Bigg(1+\frac{r}{r_s}\Bigg)^{\gamma-3} ,
\end{align}
where $\rho_s$ and $r_s$ are the scale density and scale radius, respectively. In~\cref{DMprofile}, the scale density $\rho_s$ is given by
\begin{align}
\rho_s = \rho_{\odot} \Bigg(\frac{r_{\odot}}{r_s}\Bigg)^{\gamma}\Bigg(1+\frac{r_{\odot}}{r_s}\Bigg)^{3-\gamma} ,
\end{align}
where $r_{\odot}=8.5~$kpc, $r_s=20.0~$kpc, and $\rho_{\odot}=0.4~$GeV/cm$^3$ are the distance between the Sun and the Galactic Center, scale radius, and local density, respectively~\cite{Hu:2025thq}. 
For this choice, we obtain, $D_{\rm ROI} = 3.69 \times 10^{22}~\mathrm{GeV\,cm^{-2}}$, consistent with values reported in Ref.~\cite{Cirelli:2010xx} for diffuse emission. 
Unlike annihilation signals, which scale as $\rho_{\rm DM}^2$, our CR-DM signal depends linearly on $\rho_{\rm DM}$ and is less sensitive to the choice of $\gamma$.

For the benchmark parameters in~\cref{tab:benchmarkpoints}, we compute
the differential cross section $d\sigma(E_p, E_\gamma)/dE_\gamma$ and use Eq.~\cref{eq:flux_calculation} to obtain the photon flux. We then present the results in terms of the log-binned energy flux $E_\gamma^2 d\Phi_\gamma/dE_\gamma$ across the range of $E_\gamma$.

\smallskip
\textbf{Results}---We describe the results of our analysis of CR-DM scattering, showing in~\cref{fig:moneyPlot} the predicted energy spectra of our benchmark points detailed in~\cref{tab:benchmarkpoints}. The blue solid and dashed lines represent the predicted continuous spectra for inelastic DM BP1 and BP2, respectively.  
The red solid and dashed lines in~\cref{fig:moneyPlot} are the spectra produced by the elastic DM BP1 and BP2, respectively.
The spectra from all four benchmark points appear to match the primary underlying feature of the GCE energy spectrum: a peak between 1 and 5 GeV. We show multiple benchmarks to illustrate variations in the underlying particle physics parameters that remain compatible with the data. These benchmarks demonstrate that the quality of the fit is not confined to a single choice of fine-tuned parameter, but persists across a nontrivial region of the parameter space. 

\begin{table}[]
    \centering
    \renewcommand{\arraystretch}{1.2}
    \begin{tabular}{|c|c|}
        \hline
        \textbf{Model}  & \bm{$\chi^2/$} \textbf{dof} \\
        \hline
        iDM BP1 & 36.2/10 \\
        iDM BP2 & 42.8/11 \\
        eDM BP1 & 46.6/11 \\
        eDM BP2 & 60.2/11 \\
        \hline
        DM$\to b\overline{b}$ & 46.7/12 \\
        MSPs & 73.0/13 \\
        \hline
    \end{tabular}
    \captionsetup{justification=Justified, singlelinecheck=false}
    \caption{Comparison of $\chi^2$ computed using the covariance matrix and background model in Ref.~\cite{Cholis:2021rpp} for the full sky $40\degree \times 40 \degree$ ROI, removing the inner $2\degree$ above and below the galactic plane, using the first 4 principal components of the covariance matrix, $p = 4$. The computed $\chi^2$ for DM annihilation and MSP are reproduced to within $\sim$ 5\% of the values reported in Ref.~\cite{Cholis:2021rpp}.
    }
    \label{tab:chiSq}
\end{table}

For comparison, we consider an example diffuse $\gamma$-ray emission background model described in~\cite{Cholis:2021rpp}, referred to as Cholis et al. 
The obtained GCE log-binned energy flux (black circles), along with the corresponding statistical (error bars) and correlated systematic (gray rectangles) uncertainties, is shown in~\cref{fig:moneyPlot}. 
We also show the well-known spectra of DM annihilation to $b\overline{b}$ (solid black) with best-fit parameters $m_{\rm DM}$ = 40 GeV, $\langle \sigma_Av \rangle = 1.45 \times 10^{-26} \;\rm{cm}^3/\rm{s}$ from Ref.~\cite{Cholis:2021rpp} and MSPs (dashed black) with a normalization of $6.22 \times 10^{-7} \; \rm{GeV}\,\rm{cm}^{-2} \, \rm{s}^{-1} \, \rm{sr}^{-1}$ at 1.96 GeV taken from Ref.~\cite{Cholis:2014noa}.

We use the Cholis et al. background model and their provided covariance matrix and statistical uncertainties to obtain a $\chi^2$ for our model benchmark points. We compare our model goodness of fit with the spectra produced by DM $\to b\overline{b}$ and MSPs. We use the first 4 principal components to construct the covariance matrix and compute the $\chi^2$, listing the resulting $\chi^2$ of our iDM and eDM benchmarks, as well as those of DM $\to b\overline{b}$ and MSP spectra in~\cref{tab:chiSq}. We find that, for this background model, the spectra from all iDM and eDM benchmark points yield fits\footnote{We do not perform a $\chi^2$ minimization for our models, since the background remains uncertain. We provide a $\chi^2$ for the benchmark points we consider, and minimizing over the free parameters will only reduce the $\chi^2$.} to the GCE comparable to the best-fitting spectra of DM annihilation to $b\overline{b}$ and MSPs.

There are many additional background models in the literature~\cite{DiMauro:2021raz, Abazajian:2014fta, Fermi-LAT:2015sau, Gordon:2013vta, Zhong:2019ycb, Fermi-LAT:2017opo}, producing different GCE spectra across different energy ranges and with varying bin widths. Some of these models and their corresponding statistical uncertainties are also shown (as labeled) in~\cref{fig:moneyPlot}, following Ref.~\cite{Dinsmore:2021nip, Hu:2025thq}. The goodness of fit of the CR-DM scattering, DM annihilation to $b\overline{b}$, and MSP spectra varies with the background model and its various systematic uncertainties and binning procedures as described in~\cite{Cholis:2021rpp}. 

\smallskip
\textit{Morphology---}An explanation of the GCE should produce a photon flux with both an energy spectrum and an angular distribution which are consistent with the data, once backgrounds are included.
The inferred morphology is highly sensitive to both the diffuse background model and the choice of spatial templates, as discussed in detail in Refs.~\cite{Cholis:2021rpp, Pohl:2022nnd, McDermott:2022zmq, Ramirez:2024oiw, Zhong:2024vyi, Macias:2016nev, Bartels:2017vsx, Storm:2017arh, Macias:2019omb, Ploeg:2021mrr, Pohl:2022nnd, Song:2024iup, Muru:2025vpz, Abazajian:2020tww, Hu:2026apn}. Given the strong model dependence and associated systematic uncertainties, we do not pursue a morphological analysis in this work; instead, we focus exclusively on fitting the energy spectrum.


\smallskip
\textbf{Conclusions}---
We have shown that cosmic-ray scattering with sub-GeV DM can explain the spectral features of the GeV Galactic Center excess observed by \textit{Fermi}-LAT, where the final state contains photons, in both inelastic and single-component DM scenarios.
We have found benchmark models for which the fit of the excess to the data is comparable to that of other well-studied scenarios, such as DM annihilation and MSPs, assuming the background model described in~\cite{Cholis:2021rpp}.  These benchmark models satisfy all other constraints. The observed relic abundance of DM can be achieved via thermal or non-thermal production mechanisms, or through an asymmetric dark matter scenario~\cite{Kaplan:2009ag}, without affecting our proposed GCE solutions.

To explain the GCE, one may also consider dipole interactions of the form $\overline{\chi}_2 \sigma_{\mu\nu} F^{\mu\nu} \chi_1$, 
which we will consider in a future work~\cite{us}.
Additionally, the simplified models we considered can be further explored. 

Signs of CR-DM scattering can be probed by other means, for example, in low-energy photon signatures or a higher energy $\gamma$-ray excess~\cite{Totani:2025fxx}. 
Future telescope projects like the Advanced Particle-astrophysics Telescope~\cite{APT:2021lhj} and Very Large Area gamma-ray Space Telescope~\cite{Zhang:2024jha} could provide additional measurements of GeV-energy photons at the Galactic Center and dSphs, providing further insight into the GCE and its source. 
Additional signals from these DM models could be probed by upcoming neutrino detectors such as DUNE~\cite{DUNE:2018tke, DUNE:2020ypp, DUNE:2020txw, DUNE:2020lwj, DUNE:2024qgl, Berger:2019ttc}, Hyper-K~\cite{Hyper-Kamiokande:2018ofw}, and JUNO~\cite{JUNO:2021vlw, Chauhan:2021fzu}, enabling a multi-messenger approach.

\bigskip
\section*{Acknowledgments}
We thank Aparajitha Karthikeyan, Kevin Kelly, Doojin Kim, Louis Strigari, and Douglas Tuckler for helpful discussions. JK is supported by DOE grant DESC0010504B. BD, DG, MR, and DS are 
supported by DOE Grant DESC0010813.
BD, JK, and DS wish to acknowledge the Center for Theoretical Underground Physics and Related Areas (CETUP*), the Institute for Underground Science
at Sanford Underground Research Facility (SURF), and the South Dakota Science and Technology Authority for hospitality and financial support,
as well as for providing a stimulating environment. 

\appendix

\section{\label{FormFactor}Details on the Scalar Proton Form Factor}

\noindent
We parameterize the scalar proton form factor as
\begin{equation}
    F(Q) = \frac{F(0)}{\left(1 +\frac{Q^2}{\Lambda^2}\right)},
\end{equation}
where, $\Lambda^2=0.71~$GeV$^2$~\cite{DelNobile:2021wmp} and $F(0)$ is the nucleon-scalar form factor at zero momentum transfer. It receives
contributions from quarks via
\begin{equation}
    F(0) = \sum_{q} f_q\,\frac{m_{P}}{m_q},
\end{equation}
with $m_{P}$ the proton mass, $m_q$ are the quark masses, and $f_q$ express the light quark contribution to proton mass. For this analysis, we used $F(0)=13.2$, which can be derived using the values of $f_q$ provided in Refs~\cite {Cirelli:2013ufw, DeRomeri:2024iaw, Dutta:2025fgz}. 

\bibliography{GCEExcess_DM}

\end{document}